\documentclass[prl,twocolumn,letterpaper]{revtex4}%
\usepackage{amsmath}
\usepackage{graphicx}
\usepackage{graphics}
\usepackage{amsfonts}
\usepackage{amssymb}%
\begin{document}
\title{Collective Behavior of the Closed-Shell Fermi Gas}
\author{Seth T. Rittenhouse,$^1$ M. J. Cavagnero,$^2$ Javier von Stecher,$^1$ and Chris H. Greene$^1$}
\affiliation{$^1$Department of Physics and JILA, University of Colorado, Boulder, Colorado 80309-0440\\ 
$^2$Department of Physics and Astronomy, University of Kentucky, Lexington, Kentucky 40506-0055}

\begin{abstract}
We propose an unconventional description for the ground state and collective
oscillations of the two-component normal Fermi gas with two-body zero-range
interactions. The many-body problem can be accurately reduced to a linear,
one-dimensional Schr\"{o}dinger equation in a single collective coordinate,
the hyperradius R of the $N$-atom system which is the root mean square radius.
The calculated properties of the Fermi gas ground state are shown to agree
accurately with results from the Hartree-Fock (HF) approximation over a wide
range of interspecies scattering lengths. The breathing mode excitation
frequency deviates qualitatively from HF predictions, but we show that this
reflects a failure of the HF approach for this observable.

\end{abstract}


\maketitle

Tremendous excitement has ensued in the ultracold physics community since
dilute 2-component Fermi gases have recently been shown to exhibit BCS-like
pairing physics.\cite{RegalGreinerJin2004}  This pairing occurs at sufficiently low temperature $T<<T_{f}$, sufficiently
attractive interatomic scattering length $a$, and sufficiently high density
$\rho\equiv k_{F}^{3}/(3\pi^{2})$, where $k_{F}$ is the Fermi wavenumber. This
intriguing physical regime of the BEC-BCS crossover arises in the range
$k_{F}a<-1$. The DFG currently studied in this region experimentally are in a
metastable state whose decay to a more compact geometry in a different region
of configuration space is inhibited by a multiparticle tunnelling barrier.
Comparatively little attention has been given to the behavior of the
low-temperature normal Fermi fluid with attractive interactions. In this
Letter we discuss the behavior of this normal Fermi fluid from a viewpoint
that departs from conventional mean-field theory. This theoretical framework
and subsequent improvements may prove useful for analyzing other many-body
systems, because in contrast to the familiar mean-field theories that require
the solution of inherently nonlinear differential or integro-differential
equations, the present treatment utilizes only linear Schr\"{o}dinger
equations throughout.

Our treatment is related to the hyperspherical coordinate treatments that have
been applied to finite nuclei, as in the ``K-harmonic" method reviewed by
Smirnov and Shitikova.\cite{SmirnovShitikova} The formulation is a rigorous
variational many-body calculation, aside from the usual limitations of the
assumed pairwise Hamiltonian with zero-range interactions.\cite{EsryGreene99} 
As was found in a similar study for bosons, the resulting
hyperspherical coordinate description again provides a \textit{linear},
one-dimensional, effective Schr\"odinger equation in the hyperradius. In this
initial study of the usefulness of this methodology, we restrict
ourselves to the study of fermionic systems with completely filled
shells. This is done for analytical and computational
simplification; the same approach should be equally applicable
to open-shell fermi systems, with modest extensions.

Consider $N$ identical fermions confined either magnetically or optically in a
spherically symmetric harmonic oscillator trap with angular frequency $\omega
$. Half of the fermions are in one spin substate, while the other half are in
a second spin substate. The full $N$-body time-independent Schr\"{o}dinger
equation for this system reads $H\psi=E\psi$, with
\[
H=\sum_{i=1}^{N}\left(  -\dfrac{\hbar^{2}}{2m}\nabla_{i}^{2}+\dfrac{1}%
{2}m\omega^2 r_{i}^{2}\right)  +\sum_{i>j}U_{int}\left(  \vec{r}_{i}-\vec{r}%
_{j}\right)  .
\]
Here $m$ is the mass of the fermions under consideration and $U_{int}$ is the
pairwise atomic interaction potential. We adopt the Fermi pseudopotential
$U_{int}\left(  \vec{r}_{i}-\vec{r}_{j}\right)  =\dfrac{4\pi\hbar^{2}a}%
{m}\delta\left(  \vec{r}_{i}-\vec{r}_{j}\right)  $ where $a$ is the two body
$s$ wave scattering length. Next we transform to a set of hyperspherical
coordinates. The hyperradius for this system of equal mass particles is the
root mean square distance of the atoms from the trap center, $R=\left(  \sum
r_{i}^{2}/N\right)  ^{1/2}$. The remaining $3N-1$ dimensions are described by
a set of angular coordinates collectively denoted by $\Omega$. In addition to
the ordinary set of $2N$ spherical polar coordinates for each particle
$\left\{  \left(  \theta_{i},\phi_{i}\right)  \right\}  _{i=1}^{N}$, we follow
Ref. \cite{SmirnovShitikova} in defining $N-1$ hyperangles $\alpha_{i}$,
$i=1,2,3,...,N-1.$ as:
\[
\tan\alpha_{i} \equiv (\sum_{j=1}^{i}r_{j}^{2})^{1/2}/{r_{i+1}}.
\]

In this coordinate system the kinetic energy becomes \cite{avery}%
\[
\dfrac{-\hbar^{2}}{2M}\left(  \dfrac{1}{R^{3N-1}}\dfrac{\partial}{\partial
R}R^{3N-1}\dfrac{\partial}{\partial R}-\dfrac{\mathbf{\Lambda}^{2}}{R^{2}%
}\right)
\]
where $M=Nm$ is the total mass of the system and the squared grand angular
momentum operator $\mathbf{\Lambda^{2}}$ is defined in the standard fashion
\cite{SmirnovShitikova,avery}. The trap potential is purely hyperradial, i.e.
$\sum_{i=1}^{N}\dfrac{1}{2}m\omega^2 r_{i}^{2}=\dfrac{1}{2}M \omega^2 R^{2}.$ Multiplying
$\psi$ by $R^{\left(  3N-1\right)  /2}$ allows us to eliminate first
derivative terms. One can view the rest of the calculation as the simplest
approximation to Macek's adiabatic hyperspherical description \cite{Macek1968}%
or as a one-term
truncation of an expansion of the exact $\psi\left(  R,\Omega\right)  $ into
the complete, orthonormal set of eigenfunctions of the operator
$\mathbf{\Lambda}^{2}$, multiplied by unknown radial functions that are
optimized variationally. These eigenfunctions, or \textquotedblleft
hyperspherical harmonics\textquotedblright\ (HHs, see \cite{SmirnovShitikova,avery}) can be expressed analytically in terms of Jacobi
polynomials, for any number of particles. Their eigenvalue equation is%
\[
\mathbf{\Lambda}^{2}Y_{\lambda\nu}\left(  \Omega\right)  =\lambda\left(
\lambda+3N-2\right)  Y_{\lambda\nu}\left(  \Omega\right)
\]
where $\lambda=0,1,2,..$ is the order of the harmonic. The second index, $\nu
$, stands for the further set of $N-2$ quantum numbers required to distinguish
the (usually quite large) set of degenerate states having any chosen value of
$\lambda$. This allows for the choice of a trial wave function $\psi\left(
R,\Omega\right)  =G\left(  R\right)  \Phi_{\lambda}\left(  \Omega\right)  $
where $\Phi_{\lambda}\left(  \Omega\right)  $ is a linear combination of HHs
(multiplied by appropriate spinors) with grand angular momentum $\lambda$,
which satisfies the antisymmetry requirements of this fermionic system.
Generally the coefficients of this expansion are very difficult to obtain; in
previous studies they have usually been calculated
recursively\cite{SmirnovShitikova,cavagnero,barnea} using a coefficient of
fractional parentage (cfp) expansion. To eliminate the need for a demanding
cfp calculation with hundreds or thousands of particles, we restrict our
treatment here to only magic numbers of particles, i.e., to \emph{filled
energy shells} of the noninteracting spherical harmonic oscillator. Because
the $N$-particle oscillator has a separable Schr\"{o}dinger
equation in either independent particle coordinates or hyperspherical
coordinates, any nondegenerate, normalized, antisymmetric $N$-body state must
be identical in the two coordinate systems. The nondegenerate wavefunction for
$n$ completely filled shells ($n=1,2...$) can thus be written as a Slater
determinant of the independent particle states, which allows us to construct
$\Phi_{\lambda}\left(  \Omega\right)  $ as:%
\begin{equation}
\frac{1}{\sqrt{N!}F_{\lambda}\left(  R\right)  }\sum\limits_{P}\left(
-1\right)  ^{p}P\prod\limits_{j=1}^{N}R_{n_{i}\ell_{i}}\left(  r_{i}\right)
y_{\ell_{i}m_{i}}\left(  \omega_{i}\right)  \left\vert m_{s_{i}}\right\rangle .
\end{equation}
Here $R_{n_{i}\ell_{i}}\left(  r_{i}\right)  $ is the radial solution to the
independent particle harmonic oscillator for the $i$th particle\textbf{ }given
by $rR_{n\ell}\left(  r\right)  =N_{n\ell}\exp\left(  -r^{2}/2l^{2}\right)  \left(
r/l\right)  ^{\ell+1}L_{n}^{\ell+1/2}\left[  \left(  r/l\right)  ^{2}\right]  $
where $L_{n}^{\alpha}\left(  r\right)  $ is a modified Laguerre polynomial
with $l=\sqrt{\hbar/m\omega}$. $y_{\ell_{i}m_{i}}\left(  \omega\right)  $ is an
ordinary 3D spherical harmonic with $\omega_{i}$ as the spatial solid angle
for the $i$th particle, $\left\vert m_{s_{i}}\right\rangle $ is a spin ket
that allows for two spin species of atoms, $\left\vert \uparrow\right\rangle
$and $\left\vert \downarrow\right\rangle $. $R^{\left(  3N-1\right)
/2}F_{\lambda}\left(  R\right)  =A_{\lambda}\exp\left(  -R^{2}/2L^{2}\right)
(R/L)^{\lambda+3(N-1)/2+1}$\textbf{ }is the nodeless hyperradial solution for
$N$ noninteracting particles in an oscillator with hyperangular momentum
$\lambda$ where $A_{\lambda}$ a normalization factor and $L=l/\sqrt{N}$. 
With this trial wavefunction, the expectation value of the Fermi
   pseudopotential yields an $R$-dependent effective potential
   for the function $G(R).$ From known properties of the $\delta$ function and
the standard properties of matrix elements of two-body operators between two
Slater determinantal wave functions, the desired diagonal potential matrix
elements can be shown to have the form (in harmonic oscillator units for both
energy and radius)%
\begin{equation}
\left\langle \Phi_{\lambda}\left(  \Omega\right)  |4\pi a\sum_{i>j}%
\delta\left(  \vec{r}_{i}-\vec{r}_{j}\right)  |\Phi_{\lambda}\left(
\Omega\right)  \right\rangle =\frac{4\pi aC_{\lambda}}{N^{3/2}R^{3}%
}.\label{FinalMatrixElement}%
\end{equation}
The constant $C_{\lambda}$ only depends on the
number of atoms in the system, and is found by examining the
diagonal matrix element of this hyperradial potential in the state
$F_{\lambda}\left(  R\right)  $. Integration over the hyperradius allows us to solve for $C_{\lambda}$:%
\begin{align}
C_{\lambda} &  =\dfrac{N\left(  N-1\right)  l^{5}\Gamma\left(  K+3/2\right)
}{2\Gamma\left(  K\right)  }\label{Clambda}\\
&  \times\left\langle F_{\lambda}\left(  R\right)  \Phi_{\lambda}\left(
\Omega\right)  \right\vert \delta\left(  \vec{r}_{1}-\vec{r}_{2}\right)
\left\vert F_{\lambda}\left(  R\right)  \Phi_{\lambda}\left(  \Omega\right)
\right\rangle .\nonumber
\end{align}
Here $K=\lambda+3\left(  N-1\right)  /2$ and $F_{\lambda}\left(  R\right)
\Phi_{\lambda}\left(  \Omega\right)  $ is the fully anti-symmetrized,
non-interacting atom wave function.\textbf{ }Thus the matrix element
$\left\langle F_{\lambda}\left(  R\right)  \Phi_{\lambda}\left(
\Omega\right)  \right\vert \delta\left(  \vec{r}_{1}-\vec{r}_{2}\right)
\left\vert F_{\lambda}\left(  R\right)  \Phi_{\lambda}\left(  \Omega\right)
\right\rangle $ is a sum of integrals over products of Laguerre polynomials
in independent particle coordinates.\textbf{ }Every diagonal matrix
element coefficient $C_{\lambda}$ for the $n$-th complete filled shell can be
written as a rational number multiplied by $\sqrt{2}/\pi^{2}.$ For instance,
the $n=2$ filled shell state has $N=8$ particles, and in its ground state,
grand angular momentum $\lambda=6,$ and $C_{\lambda}$ has the value
(134217728/570285)$\sqrt{2}/\pi^{2}.$ For the $n=3$ filled shell with 20
particles, $C_{\lambda}$ involves the ratio of two integers, the larger of
which is in the numerator with 34 digits. In view of the rapid increase as the number of particles is increased, we focus on the asymptotics in
the large $N$ limit. The following functional form is found to be 
reasonably accurate for all $N$ beyond about $N=20$ particles, namely
$C_{\lambda}(N) \rightarrow 0.16802 N^{7/2}\sqrt{2}/\pi^{2}.$ The fractional
error in this approximation is smaller than 1\% for $N=20$ particles, and
smaller than 0.1\% for the 30th filled shell with 9920 particles, and beyond.
The computational strategy has enabled us to carry out 
explicit calculations for as high as the 50th magic number, 
which is a degenerate fermi gas with 44,200 atoms.  This is high enough to 
extract the leading behavior in the high-$N$ limit.

This effective potential now gives a linear, one-dimensional
Schr\"{o}dinger equation in the hyperradius $h\left[  R^{\left(  3N-2\right)
/2}G\left(  R\right)  \right]  =E\left[  R^{\left(  3N-2\right)  /2}G\left(
R\right)  \right]  $ where $h$ (in oscillator units) is given by%
\begin{equation}
-\dfrac{1}{2N}\dfrac{d^{2}}{dR^{2}}+\dfrac{K\left(  K+1\right)  }{2NR^{2}%
}+\dfrac{1}{2}NR^{2}+\frac{4\pi aC_{\lambda}}{N^{3/2}R^{3}}.\label{completeH}%
\end{equation}
It is informative to recast this radial Schr\"{o}dinger equation into
dimensionless form with $E=E_{NI}E^{\prime}$ and $R=\sqrt{\left\langle
R^{2}\right\rangle _{NI}}R^{\prime}$, where $E_{NI}=\left(  \lambda
+3N/2\right)  $ and $\left\langle R^{2}\right\rangle _{NI}=\left(
3/2+\lambda/N\right)  $ are the energy and expectation values of $R^{2}$ of
the non-interacting ground state (in oscillator units). The resulting linear
effective one-dimensional Schr\"{o}dinger equation is given by%
\begin{equation}
\left(  -\dfrac{1}{2m^{\ast}}\dfrac{d^{2}}{dR^{\prime2}}+V_{eff}\left(
R^{\prime}\right)  -E^{\prime}\right)  R^{\prime\left(  3N-1\right)
/2}G\left(  R^{\prime}\right)  =0\label{rescale}%
\end{equation}
where $m^{\ast}\equiv\left(  \lambda+3N/2\right)  ^{2}$, and, in the large $N$
limit, $V_{eff}\left(  R\right)  $ is given by%
\begin{equation}
V_{eff}\left(  R^{\prime}\right)  =\dfrac{1}{2R^{\prime2}}+\dfrac{1}%
{2}R^{\prime2}+\dfrac{4\pi a C_{\lambda}}{l\left(  \lambda
+3N/2\right)  ^{5/2}R^{\prime3}}.\nonumber
\end{equation} 
The value of $\lambda$ for each filled shell ground state can be ascertained by
comparing the energy of $N$ noninteracting particles in a harmonic oscillator
potential, written in each coordinate system. In the independent atom
representation the energy is given by $E=\hbar\omega\left[  \sum_{i=1}%
^{N}\left(  l_{i}+2n_{i}\right)  +3N/2\right]  $. The energy of $N$
noninteracting particles, described using hyperspherical quantum numbers, with
grand angular momentum $\lambda$ and $\chi=0,1,2,...$ hyperradial nodes in
$F_{\lambda}\left(  R\right)  $, is given by $E=\hbar\omega\left(
\lambda+2\chi+\dfrac{3N}{2}\right)  .$ The ground state has
$\chi=0,$ while the first breathing mode excited state has $\chi=1$. After we
equate the two expressions for the ground state energy, we obtain
$\lambda=\sum_{i=1}^{N}\left(  l_{i}+2n_{i}\right) $. 
In terms of the number of complete filled shells, $n$,
namely $N(n)=n(n+1)(n+2)/3,$ and $\lambda(n)=(n-1)n(n+1)(n+2)/4$.  To a good
approximation at all $N$, $\lambda(N)\simeq(3N)^{4/3}/4-3N/2+N^{2/3}%
/192^{1/3}$, which becomes exact in the limit $N\rightarrow\infty$. Similarly,
the Fermi wave number, $k_{f}$, may be shown to be $k_{f}=\sqrt{2\left(
n+1/2\right)  }/l\rightarrow\sqrt{2}\left(  3^{1/6}\right)  N^{1/6}/l$.
This all shows that in the large-$N$ limit%
\[
V_{eff}\left(  R^{\prime}\right)  \rightarrow\dfrac{1}{2R^{\prime2}}+\dfrac
{1}{2}R^{\prime2}+\zeta\dfrac{k_{f}a}{R^{\prime3}},
\]
where $\zeta\simeq0.16802\left(  64\sqrt{12}/81\pi\right).$ 
Note that the only parameter in this asymptotic large-$N$ potential
curve is in fact $k_{f}a$. \ At  $N\rightarrow\infty$ we see that
$m^{\ast}\rightarrow N^{8/3}$, whereby the second derivative term in
equation \ref{completeH} is negligible. \ This means the squared zero-order ground
state wave function in the large $N$ limit is approximated by $\left[ R^{\prime
\left(  3N-1\right)  /2}G\left(  R^{\prime}\right) \right]^2  \approx\delta\left(
R^{\prime}-R_{\min}^{\prime}\right)  $ where $R_{\min}^{\prime}$\ is the
location of the minimum of $V_{eff}\left(  R^{\prime}\right)  $, if it exists.
This recognition of the strong hyperradial localization 
for infinite dimensional systems is also used extensively 
in dimensional perturbation theory.\cite{GoodsonHerschbach}
Ground state expectation values of quantities 
involving the hyperradius can be extracted by simply evaluating them at
$R_{\min}^{\prime}$ (e.g. $\left\langle R^{\prime2}\right\rangle =R_{\min
}^{\prime2}$ and $E=V_{eff}\left(  R_{\min}^{\prime}\right)  $).

\begin{figure}[h]
\begin{center}
\includegraphics[width=3in]{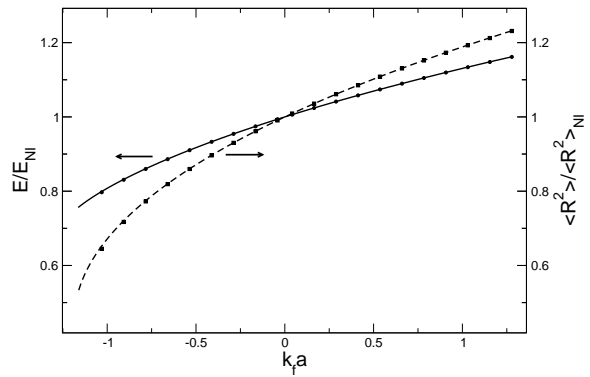}
\end{center}
\caption{The ground state energy (solid line) and $\left\langle R^{2}%
\right\rangle $ (dashed line) calculated using the hyperspherical method are
compared with the Hartree-Fock results (circles and squares,
respectively), as functions of $k_{f}a$. \ All calculations on this figure were carried out for 240 atoms.}%
\label{RandEfig}%
\end{figure}

The effective potential in equation \ref{completeH} is similar in form to the
analogous hyperradial potential obtained for the Bose-Einstein condensate (BEC) in Ref.
\cite{bohn_esry_greene}. The main differences between these two potentials
stem from the antisymmetry requirements; whereas the constant in the interaction potential
\cite{bohn_esry_greene} acted as $N^{3/2}$ for a BEC, we find for the DFG an
$N^{7/2}$ behavior for $N>>1$. On the other hand, the
coefficient of the $R^{-2}$ effective centrifugal repulsion increases as
$N^{5/3},$ as opposed to $N$ for the BEC case, which is a manifestation of the
Fermi pressure in the DFG. The hyperspherical results 
for the ground state energy and expectation
value of $R^{\prime2}$ are very close to the HF
predictions, for both repulsive and attractive interactions, 
as is evident from Fig. \ref{RandEfig}.
\vskip 0.2in

The effective potential energy $V_{eff}$ for values of $k_{f}a\geq0$, i.e. for
repulsive interactions $a>0$, shows only one global minimum in $V_{eff}\left(
R^{\prime}\right)  $, and the condensate energy increases 
as does the overall size of the DFG. \ For values of $k_{f}a<0$, 
where the interspecies interaction
is attractive, the energy of the ground
state decreases below the non-interacting energy as the atoms are pulled in
together by the attractive interaction and by the trap. Provided
$ k_{f}a $ is not too negative, the repulsive $1/R^{2}$
kinetic energy term is sufficiently large and positive to stabilize the gas in
a local hyperradial potential minimum. But when $k_{f}a$ decreases to
sufficiently negative values, the local minimum in $V_{eff}\left(  R\right)  $
disappears altogether at a critical value we denote $a_{c}$, 
given in the large $N$ limit by $k_{f}a_{c}\simeq-1.21$. This prediction is
in good agreement with the value of the critical scattering length found by
the authors of Ref. \cite{bruun_burnett} beyond which there was no
self-consistent Hartree-Fock solution. This value for the critical scattering
length differs by less than $8\%$ from the value of $aN^{1/6}/l=-.66$ found by
the authors of \cite{houbiers}. Thus we see that there are two regions of
interest $k_{f}a_{c}<k_{f}a<0$ and $k_{f}a\leq k_{f}a_{c}$. 

\begin{figure}[h]
\begin{center}
\includegraphics[width=3in]{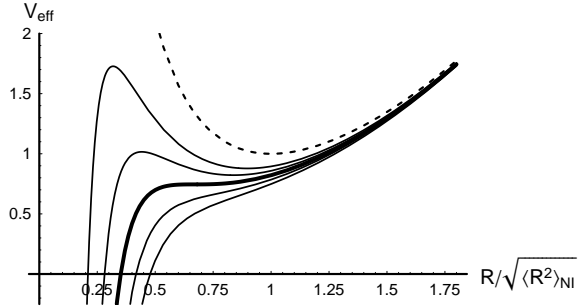}
\end{center}
\caption{The effective potential for five values of $k_{f}a$ spaced evenly
from $k_{f}a_{c}-0.5$ to $k_{f}a_{c}+0.5$. The bold curve is at $k_{f}%
a=k_{f}a_{c}$, and the dotted curve is the non-interacting potential.}%
\label{Veff}%
\end{figure}

Figure \ref{Veff}
shows $V_{eff}\left(  R^{\prime}\right)  $ for five equally spaced values
of $k_{f}a$ about $k_{f}a_{c}$ ranging over $k_{f}a_{c}-0.5$ to $k_{f}%
a_{c}+0.5$, in steps of 0.25. The behavior of $V_{eff}$ here resembles the
effective hyperradial potential for N identical bosons. BECs formed with $a<a_c$ were predicted to collapse by a number of
studies \cite{Shuryak96,Stoof97,bohn_esry_greene},
and the corresponding phenomenon of the bosenova was observed
experimentally.\cite{Bradley95,Donely} 

\begin{figure}[h]
\begin{center}
\includegraphics[width=3in]{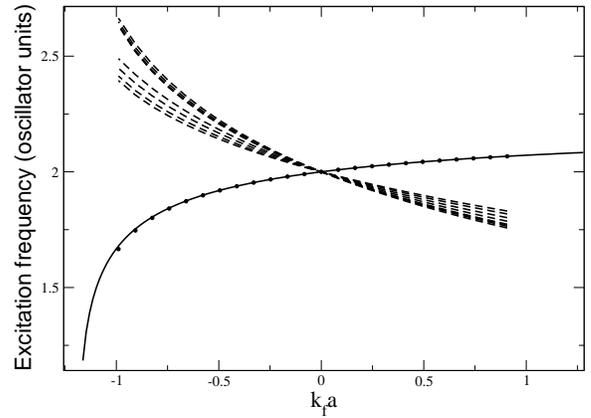}
\end{center}
\caption{The lowest excitation energy is shown as a function of $k_{f}a$. The
negatively sloping dashed lines are the lowest 8 frequencies
predicted by Hartree-Fock. The positively sloping solid line is the
hyperspherical prediction and the circles are the predictions from the sum
rule calculation. All of these calculations were
carried out for the closed shell system having a total of 240 atoms.}%
\label{frequency}%
\end{figure}

Moreover, as $k_{f}a$ is lowered, the local minimum of $V_{eff}\left(
R^{\prime}\right)  $ becomes \textquotedblleft softer", 
as the curvature at the local minimum decreases. 
This suggests that the lowest excitation frequency
(the frequency of the breathing mode) should decrease as $a$ is made
increasingly negative. \ Figure \ref{frequency} compares the excitation frequency
calculated in our hyperspherical method with the lowest one
predicted using the sum rule method,\cite{Vichi} 
in which the desired moments were calculated using the 
Hartree-Fock ground state. The calculated HF 
frequencies for the lowest 8 excitations caused by the same $R^{2}$
breathing mode operator are also shown.\ The hyperspherical 
frequencies are in excellent agreement
with the sum rule predictions, but the Hartree-Fock 
frequencies are qualitatively
different and apparently erroneous. \ 
Our hyperspherical description allows for the
calculation of \emph{collective} monopole excitations in which every particle
in the cloud oscillates in and out at the characteristic frequency, while
Hartree-Fock predicts only single-particle excitations in which just one
orbital of the DFG gets excited. 

In summary, we have developed a fully antisymmetrized,
variational description of $N$ fermionic atoms, with zero-range interactions
in a spherical harmonic trap. \ Truncation to the lowest antisymmetric
hyperspherical harmonic reduces the problem to a linear, one-dimensional
Schr\"{o}dinger equation in the hyperradius with an effective potential. 
In the large atom number limit, it only depends on the quantity $k_{f}%
a$.\ The ground state energy and radius predicted from this linear quantum
mechanical treatment map accurately onto those predicted with the Hartree-Fock
method, while the breathing mode frequencies agree with the sum rule and are a
qualitative improvement over the calculated HF frequencies. This picture
predicts a ground state energy that collapses to $-\infty$ at $k_{f}a<-1.21$,
in a manner identical to the physics of the "bosenova". \cite{bohn_esry_greene}%
 But the full interrelationship between this phenomenon and the physics of
pairing in the BEC-BCS crossover region goes beyond the scope of the present
study and will be explored in a subsequent publication.

This work was supported in part by the National Science Foundation.

\end{document}